\documentclass[osajnl,twocolumn,preprint, superscriptaddress,10pt]{revtex4-1} 
\usepackage{amsmath,amssymb,graphicx}

\begin{document}
\title{Impact of cavity spectrum on span in microresonator frequency combs}
\author{Ivan S. Grudinin}
\email{Corresponding author: grudinin@jpl.nasa.gov}
\affiliation{Jet Propulsion Laboratory, California Institute of Technology, 4800 Oak Grove dr., Pasadena, California 91109, USA}
\author{Lukas Baumgartel}
\affiliation{Jet Propulsion Laboratory, California Institute of Technology, 4800 Oak Grove dr., Pasadena, California 91109, USA}
\affiliation{Physics and Astronomy, University of Southern California, Los Angeles, California 90089, USA}
\author{Nan Yu}
\affiliation{Jet Propulsion Laboratory, California Institute of Technology, 4800 Oak Grove dr., Pasadena, California 91109, USA}
\begin{abstract}We experimentally study the factors that limit the span in frequency combs derived from the crystalline whispering gallery mode resonators. We observe that cavity dispersion is the key property that governs the parameters of the combs resulting from cascaded four wave mixing process. Two different regimes of comb generation are observed depending on the precise cavity dispersion behavior at the pump wavelength. In addition, the comb generation efficiency is found to be affected by the crossing of modes of different families. The influence of Raman lasing is discussed.
\end{abstract}
\maketitle 
\noindent  Femtosecond frequency combs have revolutionized precision measurement, optical clocks,
communications and spectroscopy. While there have been notable breakthroughs, several years after the
demonstrations of the first micro-comb based on miniature whispering gallery mode (WGM) resonators \cite{firstcomb, grudinin-caf2, savchenkov-caf2} many questions remain open, preventing realization of its full potential \cite{sciencereview}. For example, understanding of the mode locking mechanisms,  stability of the RF beat note signals, and relationship between the comb span and cavity parameters is only beginning to emerge \cite{tsolitons, universal, silica, vahaladisk, chaoticdynamics, gaetaroute,yu1, yu2, yu3, erkintalo}. While various aspects of the comb generation have been studied, the comb span limitations have not been explored in detail. A nearly octave comb span and mode locking will be required for optical clocks and optical metrology. Comb spacing also needs to be small enough for electronic detection. It was observed that the comb span generally grows with pumping power, however the coherence of the beatnote is always lost in a broad comb due to generation of the sub--combs \cite{universal, nitrideoctave, silicaoctave}. Moreover, observed comb spans do not always increase with higher power and the causes of this bandwidth limitation are not well understood. In this paper we report results of our experimental investigation of how the pump wavelength and dispersion, mode crossing and Raman lasing influence the comb span and generation efficiency. Our findings suggest a path towards the realization of an octave spanning, coherent microcomb.

To investigate factors limiting the span of micro--combs we have fabricated a number of MgF$_2$ WGM resonators of various size and shape. Each resonator has its axis aligned with the crystalline optical axis (z--cut). We use analytical approximations and finite element method (FEM) \cite{freefem} to analyse spectrum and dispersion of our resonators. The resonators are used to generate combs for the experimental part of the study.

\begin{figure}[htb]
\centerline{\includegraphics[width=5.0cm]{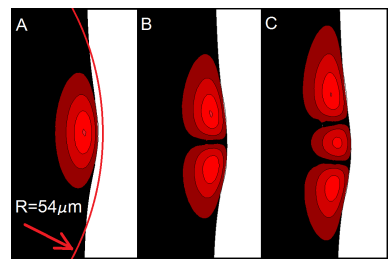}}
\caption{\label{fig:modes} Resonator geometry and the first three mode families. The shown area is 50 $\mu m$ high and 75 $\mu m$ wide. A) l-m=0, fundamental mode along with a segment of a circle with 54 $\mu m$ radius, B) l-m=1 mode, c)l-m=2 mode. (Color online)}
\end{figure}

To study the influence of dispersion and mode crossings on comb span we fabricated a resonator supporting only a few families of modes \cite{singlemode}. We used FEM \cite{sensor} to calculate the free spectral range (FSR) of the first three families of modes as shown in Fig. \ref{fig:modes}. The resonator profile image was taken through a microscope and digitally processed to extract its shape. The experimentally measured FSR of the combs produced by this resonator is $F$=46.027$\pm$0.003 GHz. During the FEM analysis the cavity radius was changed to fit the fundamental FSR to $F$. Our FEM solutions used adaptive meshing with over 30k elements, providing absolute frequency precision of a few MHz. It was found that the FSR of the l-m=1 and l-m=2 mode families are 0.02 \% and 0.06\% larger than the fundamental mode FSR. This means that mode crossings in these basic families of WGMs are possible in the comb pump wavelength region. Optical image processing in Fig. \ref{fig:modes} is not accurate enough to determine the frequencies of the non--fundamental modes precisely. The error stems from determining the exact resonator boundary shape. In addition, these families may be shifted in a complex way during the comb generation due to partial overlap with the fundamental modes. These are the reasons why we did not attempt to identify exactly the crossing modes in this case.

\begin{figure}[htb]
\centerline{\includegraphics[width=8.4cm]{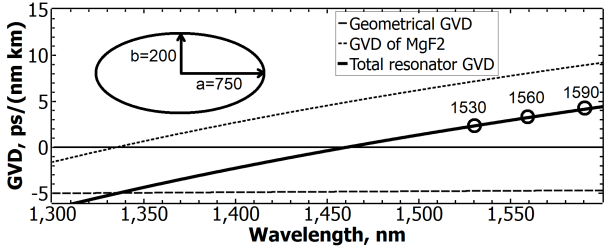}}
\caption{\label{fig:dispersion} Dispersion of the ellipsoidal resonator. Points indicate wavelength at which the frequency comb was experimentally pumped in this work.}
\end{figure}

To study the dispersion the resonator is modelled as an oblate ellipsoid, for which we can compute the spectrum using known approximations \cite{geometric}, iteratively taking the wavelength dependence of refractive index into account. The minor radius of curvature of our resonator r=54 $\mu m$ is chosen as shown in Fig. \ref{fig:modes}. Here the major ellipsoid axis is a=750 $\mu$m and the minor axis can be approximated as b=$\sqrt{ar}$=200 $\mu$m. The computed effective group velocity dispersion is shown in Fig. \ref{fig:dispersion}. To estimate the precision of this approach we note that changing b from 200 to 150 $\mu$m leads to zero dispersion wavelength (ZDW) shift from 1.459 to 1.458 $\mu$m.

To investigate the role of cavity dispersion in comb dynamics we pumped the fundamental modes of this resonator at 1530 nm, 1560 nm and 1590 nm wavelengths. A fiber laser with 20 kHz linewidth was used at 1560 nm while others are distributed feedback (DFB) semiconductor lasers having larger linewidths. Thermal locking was used to pump the resonator and the stable spectra were recorded. The DFB lasers used for 1530 nm and 1590 nm pumping have linewidth of around 1 MHz. Since laser linewidth exceeded cavity linewidth in these cases, the thermal lock didn't work well and we obtained the comb spectra by repeatedly scanning the lasers around the cavity mode. The comb spectra obtained this way for the case of 1560 nm pump reproduce the envelope of the comb obtained with the thermal lock, although some comb lines are missing.
\begin{figure}[htb]
\centerline{\includegraphics[width=8.4cm]{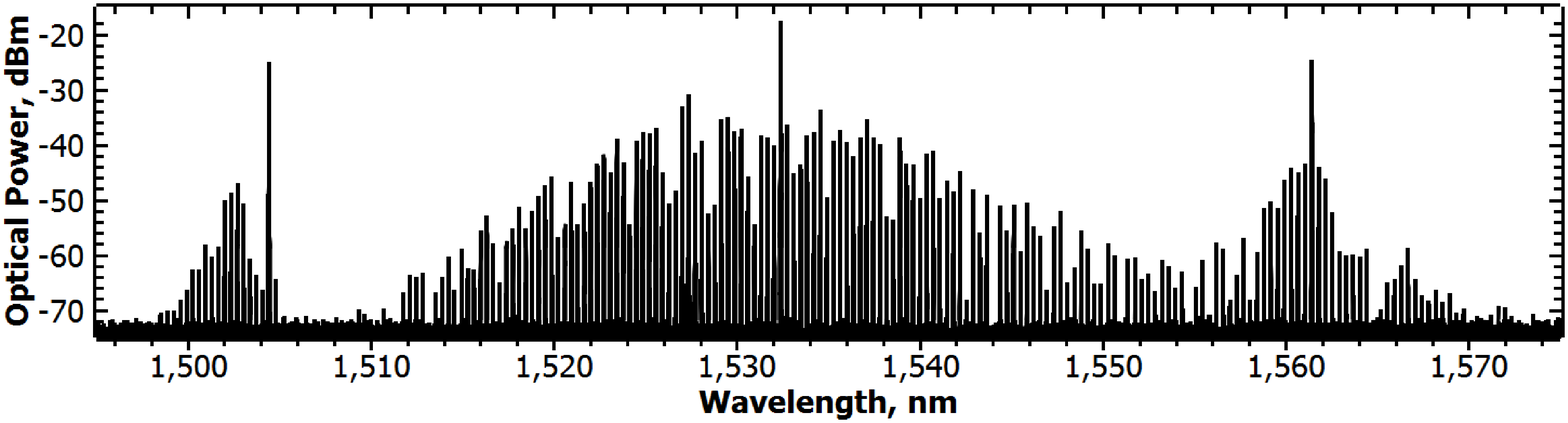}}
\centerline{\includegraphics[width=8.4cm]{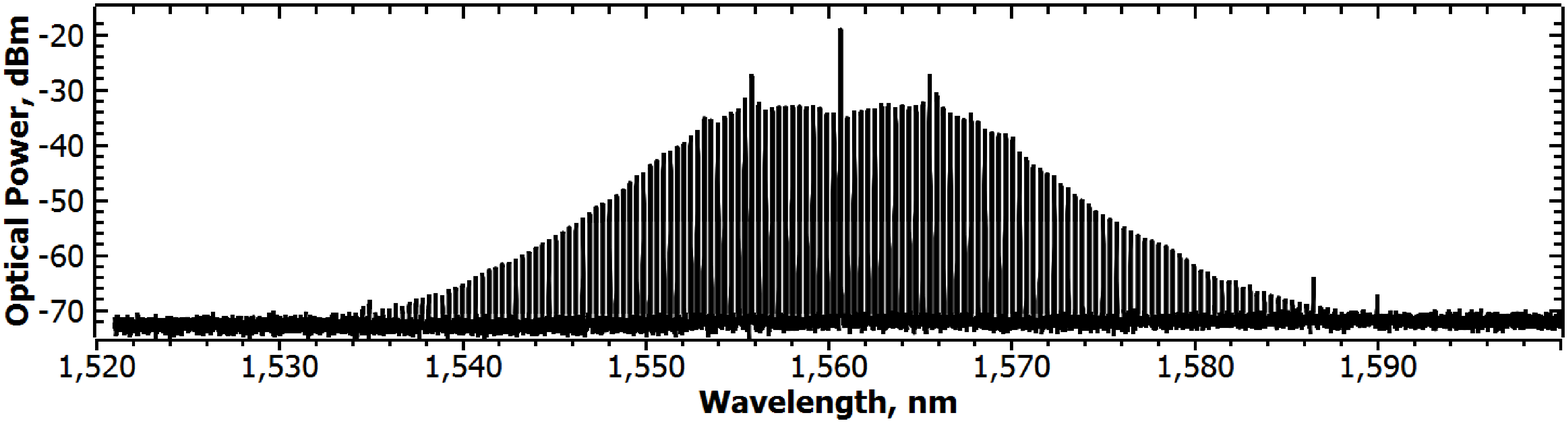}}
\centerline{\includegraphics[width=8.4cm]{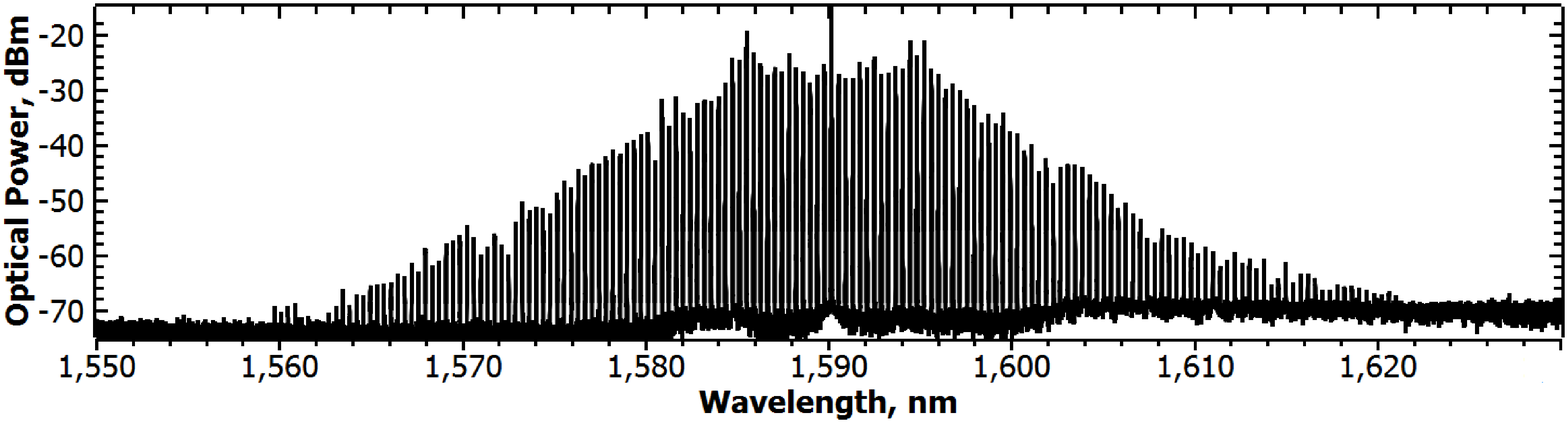}}
\caption{\label{fig:combs} The combs from 1530 (top) 1560 (middle) and 1590 (bottom) nm pump. Pump power is 9 mW.}
\end{figure}

The comb pumped at 1590 nm  started with N=12 FSR primary comb, which evolved into a comb shown in Fig. \ref{fig:combs} as the laser was tuned into the resonance. The comb pumped at 1560 nm started at N=14 FSR, and evolved similarly to the 1590 nm comb. The important observation is that a comb changes dynamics notably as the pump wavelength approached the ZDW. The 1530 nm comb did not start along the primary--secondary path. Close to threshold pump power it showed a number of sidebands spaced by 1 FSR with random amplitudes. As the laser pump was increased, a mixture of 1 FSR and 39 FSR peaks was present. Eventually at 9 mW pump power the two strong lines at 79 FSR (80 FSR at 100mW pump) developed as shown in Fig. \ref{fig:combs}. We note that the increase in N with decreasing cavity dispersion is consistent with the theoretical model for the comb excited in negative GVD regime \cite{yu1, yu2, yu3,universal}. However, the dynamics of a comb pumped near zero cavity GVD have not been extensively analyzed and is expected to be a transition towards the normal GVD combs \cite{grudinin-caf2, savchenkov-caf2, normalgvdcomb}. 

\begin{figure}[htb]
\centerline{\includegraphics[width=8.4cm]{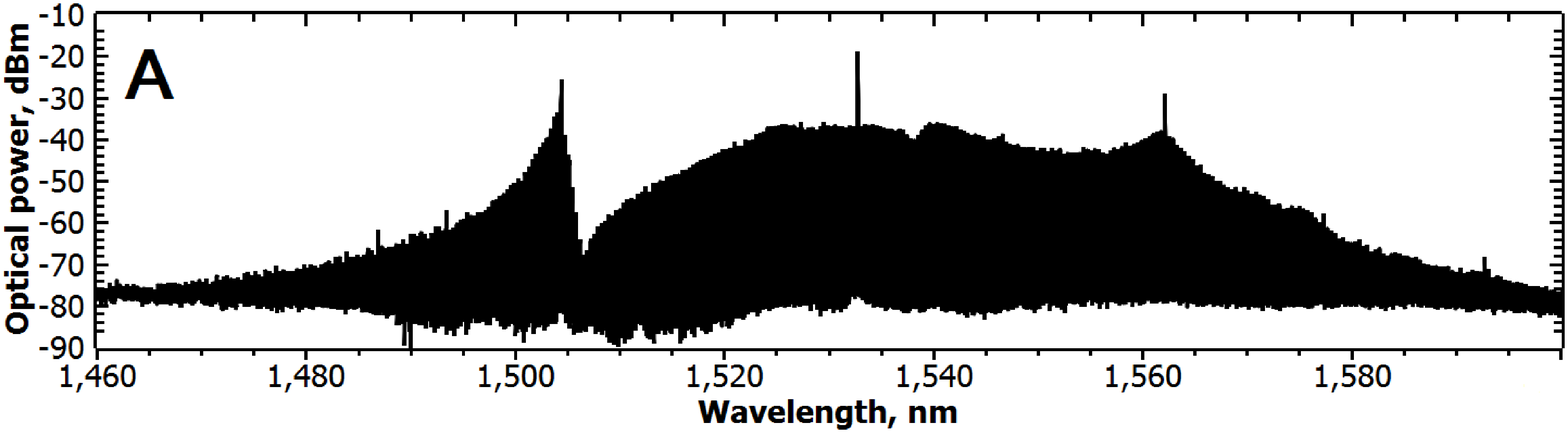}}
\centerline{\includegraphics[width=8.4cm]{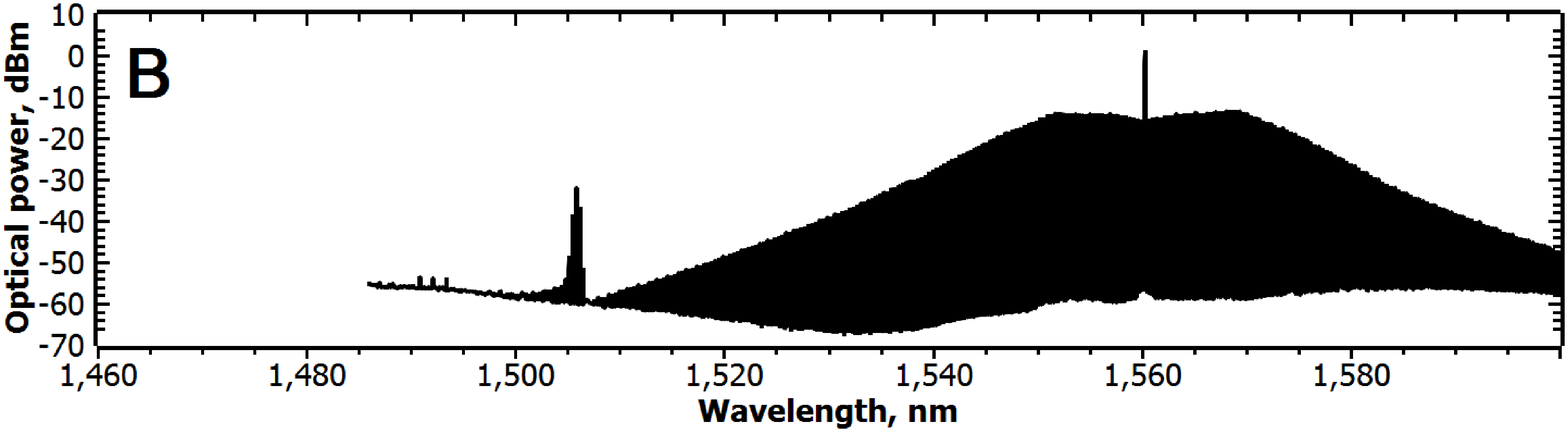}}
\centerline{\includegraphics[width=8.4cm]{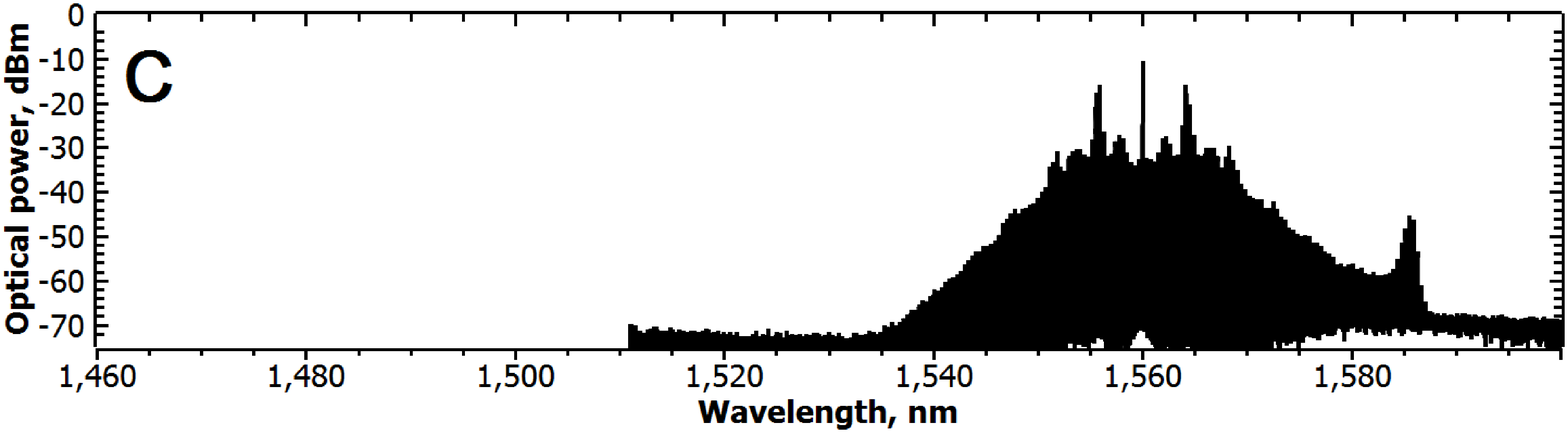}}
\caption{\label{fig:crossing} Mode crossing effect on the comb envelope. The combs shown in ``A'' and ``B'' are derived from the fundamental mode family pumped at different wavelengths. The mode crossing signature is at the same wavelength (1500 nm). The comb shown in ``C'' is derived from non-fundamental mode family and the mode crossing is at a different wavelength (1590 nm). Pump power is 100 mW.}
\end{figure}

As shown in Fig. \ref{fig:crossing}, there are local variations in FWM efficiency at around 1505 nm and 1585 nm. These spectral features may be explained by the coupling of modes from two different WGM families. The mode crossing induces local change of dispersion and thus phase matching for the FWM process, leading to increase or decrease in comb amplitude. While we have not directly confirmed that the mode crossing is present in our case, several findings point to this explanation of the spectral features. First, the FEM provides the evidence that the modes indeed cross at some wavelength in our resonator. Second, the combs in Fig. \ref{fig:crossing} A and B are centered around their corresponding pump wavelength, while the spectral features remain at a constant wavelength, indicating the intrinsic property of the cavity under investigation. Third, the spectral features in the comb excited in the fundamental family of WGMs and the features in the comb of the non-fundamental family of WGMs have different wavelength (compare Fig. \ref{fig:crossing} A-B and C).

We have also carried out experiments aimed at extending the comb span by increasing the pump power. However, we found that the comb span only weakly changes with the pump power. For example, the span of a 45 GHz repetition rate comb in a MgF$_2$ cavity increased 50\% with the 4-fold increase in the pump power from 100 mW to 400 mW at 1560 nm wavelength. 

Mode crossing is known to shift the frequencies of the modes locally, thus affecting dispersion \cite{diodespectroscopy}. The apparent improvement of comb generation efficiency near the mode crossing feature in Fig.\ref{fig:crossing}-B and -A means that the resulting dispersion change strongly influences the FWM process. Along these lines we can attribute the gradual decrease of comb generation efficiency away from the pump wavelength to the change in dispersion as shown in Fig. \ref{fig:dispersion}. The mode crossing, while not affecting the comb span directly, provides an important hint: the cavity dispersion (spectrum) needs to be engineered to generate a comb with a broader span. This idea is supported by the significant increase of span when the dispersion is flattened in other nonlinear FWM--based supercontinuum and comb generation examples \cite{nitrideoctave,supercontinuum}.

\begin{figure}[htb]
\centerline{\includegraphics[width=8.4cm]{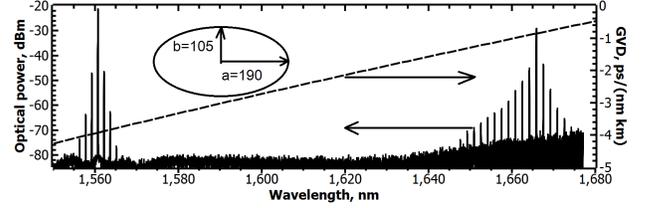}}
\caption{\label{raman} Generation of a Raman comb with normal GVD at the pump wavelength in a MgF$_2$ microresonator. The GVD of an ellipsoid approximating the cavity is shown.}
\end{figure}

It is also known that parametric gain is slightly higher in the perfectly phase matched regime than the Raman gain, which does not require phase matching \cite{kipraman}. We found that in larger MgF$_2$ resonators only parametric generation of comb is observed, while in smaller resonators the Raman lasing notably competes with the generation of combs. To study the concurrent FWM and Raman lasing experimentally, we fabricated a MgF$_2$ microcavity with FSR comparable to Raman gain linewidth, then changed the cavity temperature to tune the relative offset of the Raman line and mode positions. This cavity is nearly single mode and its dispersion can be modelled with an ellipsoid having the major axis a=190 $\mu$m and minor axis b=105 $\mu$m. The GVD of such an ellipsoid is shown in Fig. \ref{raman}.

\begin{figure}[htb]
\centerline{\includegraphics[width=8.4cm]{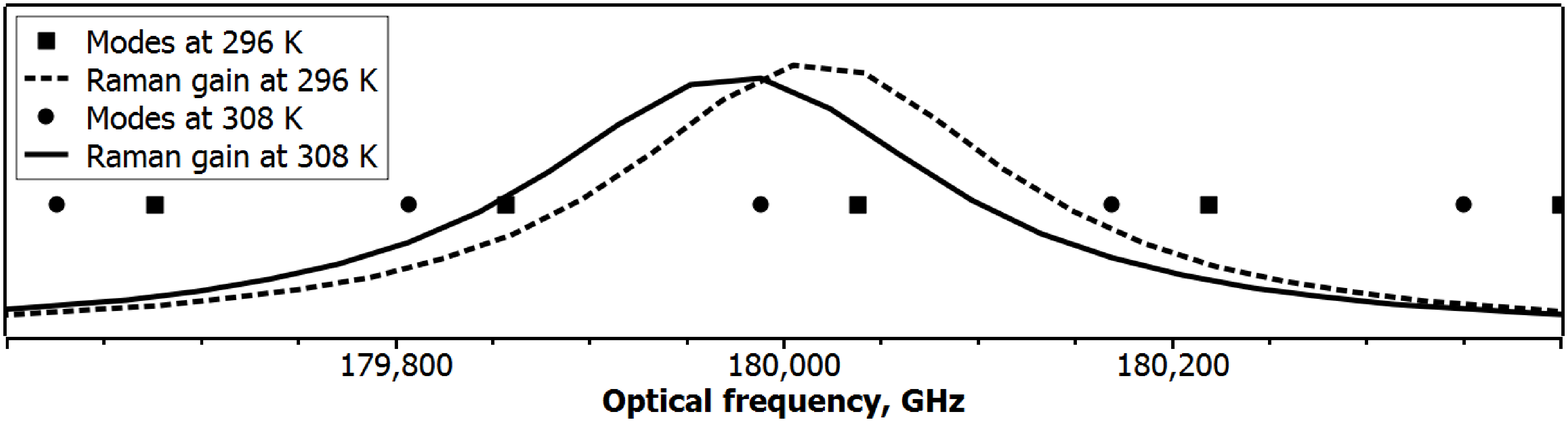}}
\caption{\label{fig:ramanpeak} Raman gain and WGM frequencies at two temperatures.}
\end{figure}

The FSR of around 180 GHz is derived from the comb line spacing. The Raman gain for the strongest phonon mode in MgF$_2$ at temperatures 296 and 307.7 K is a lorentzian having the linewidth of 241 and 253 GHz and offset from the pump of 12.148 and 12.139 THz correspondingly \cite{ramanmgf2}. The shift of the cavity modes upon heating was measured to be -50$\pm$8 GHz and the change in FSR was negligible. Thus between these two temperatures the mode closest to the Raman gain peak shifts by around 10 GHz relative to the peak and thus experiences different Raman gain at the two temperatures as shown in Fig. \ref{fig:ramanpeak}.
The exact gain is not known due to the error in FSR measurements resulting in the uncertainty (17 GHz) of the frequency of the mode near the Raman peak. The change in gain was enough, however, to observe both Raman lasing and comb generation at room temperature and only comb generation at the elevated temperature. Thus, we find that by changing the cavity temperature we can adjust the relative strengths of these two competing nonlinear processes, benefiting comb generation.

We also note that the larger resonators have anomalous GVD at pump, comb and Raman stokes wavelengths, while the smaller resonators have normal GVD for the pump and the comb wavelengths and near zero GVD for the Raman wavelength (compare Fig. \ref{fig:dispersion} and \ref{raman}). Due to this behavior of dispersion in small resonators, we were able to observe a mode--locked Raman laser as shown in Fig. \ref{raman}. This comb is observed in the regime when only the Raman lasing is initially present and is similar to the results of Ref. \cite{modelockraman}.

Thus, parametric gain always prevails in larger resonators where mode spectrum is dense enough. In smaller resonators the GVD becomes normal due to geometric contribution and the usual FWM leading to combs can not be phase matched. We attribute the comb observed in smaller resonators \cite{smcomb} and the combs observed in near zero GVD pump regime (Fig. \ref{fig:combs}-top) to a different FWM generation mechanism similar to that observed in CaF$_2$ resonators also in the normal GVD regime \cite{grudinin-caf2, savchenkov-caf2}.

In conclusion, we have investigated several mechanisms which play an important role in the span of microcombs generated in crystalline cavities. Our findings suggest that total cavity dispersion -- as well as its wavelength dependence -- determine the comb span. We observe that while mode crossings may not directly limit comb width, the resulting local changes in dispersion may increase or decrease FWM efficiency in that region. We show that by tuning cavity temperature, we may suppress the competing process of stimulated Raman scattering. Our observations suggest a route towards the broad coherent comb in crystalline WGM resonators: the spectrum must be engineered to have a flat anomalous GVD in the desired spectral region and a few if any mode crossings. The flatness of the GVD is equivalent to low third order dispersion. 
\\

This work was carried out at the Jet Propulsion Laboratory, California Institute of Technology under a contract with the National Aeronautics and Space Administration.


\begin{thebibliography}{0}%
\makeatletter
\providecommand \@ifxundefined [1]{%
 \@ifx{#1\undefined}
}%
\providecommand \@ifnum [1]{%
 \ifnum #1\expandafter \@firstoftwo
 \else \expandafter \@secondoftwo
 \fi
}%
\providecommand \@ifx [1]{%
 \ifx #1\expandafter \@firstoftwo
 \else \expandafter \@secondoftwo
 \fi
}%
\providecommand \natexlab [1]{#1}%
\providecommand \enquote  [1]{``#1''}%
\providecommand \bibnamefont  [1]{#1}%
\providecommand \bibfnamefont [1]{#1}%
\providecommand \citenamefont [1]{#1}%
\providecommand \href@noop [0]{\@secondoftwo}%
\providecommand \href [0]{\begingroup \@sanitize@url \@href}%
\providecommand \@href[1]{\@@startlink{#1}\@@href}%
\providecommand \@@href[1]{\endgroup#1\@@endlink}%
\providecommand \@sanitize@url [0]{\catcode `\\12\catcode `\$12\catcode
  `\&12\catcode `\#12\catcode `\^12\catcode `\_12\catcode `\%12\relax}%
\providecommand \@@startlink[1]{}%
\providecommand \@@endlink[0]{}%
\providecommand \url  [0]{\begingroup\@sanitize@url \@url }%
\providecommand \@url [1]{\endgroup\@href {#1}{\urlprefix }}%
\providecommand \urlprefix  [0]{URL }%
\providecommand \Eprint [0]{\href }%
\providecommand \doibase [0]{http://dx.doi.org/}%
\providecommand \selectlanguage [0]{\@gobble}%
\providecommand \bibinfo  [0]{\@secondoftwo}%
\providecommand \bibfield  [0]{\@secondoftwo}%
\providecommand \translation [1]{[#1]}%
\providecommand \BibitemOpen [0]{}%
\providecommand \bibitemStop [0]{}%
\providecommand \bibitemNoStop [0]{.\EOS\space}%
\providecommand \EOS [0]{\spacefactor3000\relax}%
\providecommand \BibitemShut  [1]{\csname bibitem#1\endcsname}%
\let\auto@bib@innerbib\@empty
\end{thebibliography}%


\begin{thebibliography}{99}
\bibitem{firstcomb} P. Del’Haye, A. Schliesser, O. Arcizet, T. Wilken, R. Holzwarth, and T. J. Kippenberg, ``Optical frequency comb generation from a monolithic microresonator,'' Nature \textbf{450}, 1214–-1217 (2007).
\bibitem{grudinin-caf2} I. S. Grudinin, N. Yu, L. Maleki, ``Generation of optical frequency combs with a CaF$_2$ resonator,'' Opt. Lett. \textbf{34}, 878 (2009).
\bibitem{savchenkov-caf2}A. A. Savchenkov, A. B. Matsko, V. S. Ilchenko, I. Solomatine, D. Seidel, and L. Maleki ``Tunable optical frequency comb with a crystalline whispering gallery mode resonator,'' Phys. Rev. Lett. \textbf{101}, 093902 (2008).
\bibitem{sciencereview} T. J. Kippenberg, R. Holzwarth, and S. A. Diddams, ``Microresonator--based optical frequency combs,'' Science \textbf{332}, 555 (2011).
\bibitem{tsolitons} T. Herr, V. Brasch, J. D. Jost, C. Y. Wang, N. M. Kondratiev, M. L. Gorodetsky, T. J. Kippenberg, ``Mode-locking in an optical microresonator via soliton formation,''
arXiv:1211.0733 .
\bibitem {universal} T. Herr, K. Hartinger, J. Riemensberger, C. Y. Wang, E. Gavartin, R. Holzwarth, M. L. Gorodetsky, and T. J. Kippenberg, ``Universal formation dynamics and noise of Kerr-frequency combs in microresonators,'' Nat. Photonics \textbf{6}, 480--487 (2012).
\bibitem{silica} S. B. Papp, S. A. Diddams, ``Spectral and temporal characterization of a fused-quartz-microresonator optical frequency comb,'' Phys. Rev. A \textbf{84}, 053833 (2011).
\bibitem{vahaladisk} J. Li, H. Lee, T. Chen and K. J. Vahala, ``Low--Pump--Power, Low--Phase--Noise, and Microwave to Millimeter--Wave Repetition Rate Operation in Microcombs,'' Phys. Rev. Lett. \textbf{109}, 233901 (2012).
\bibitem{chaoticdynamics} A. B. Matsko, W. Liang, A. A. Savchenkov and L. Maleki, ``Chaotic dynamics of frequency combs generated with continuously pumped nonlinear microresonators,'' Opt. Lett. \textbf{38}, 525-527 (2013).
\bibitem{gaetaroute} M. R. E. Lamont, Y. Okawachi, and A. L. Gaeta, ``Route to stabilized ultrabroadband microresonator-based frequency combs,'' arXiv:1305.4921.


\bibitem{yu1} Y. K. Chembo and N. Yu, ``Modal expansion approach to optical-frequency-comb generation with monolithic whispering-gallery-mode resonators,''  Phys. Rev. A  \textbf{82}, 033801 (2010).

\bibitem{yu2} Y. K. Chembo and N. Yu, ``On the generation of octave-spanning optical frequency combs using monolithic whispering-gallery-mode microresonators,'' Opt. Lett.  \textbf{35}, 2696--2698 (2010).

\bibitem{yu3} Y. K. Chembo, D. V. Strekalov, and N. Yu, ``Spectrum and Dynamics of Optical Frequency Combs Generated with Monolithic Whispering Gallery Mode Resonators,'' Phys. Rev. Lett. \textbf{104}, 103902 (2010).


\bibitem{erkintalo} S. Coen, H. G. Randle, T. Sylvestre, and M. Erkintalo, ``Modeling of octave-spanning Kerr frequency combs using a generalized mean-field Lugiato–Lefever model,'' Opt. Lett. \textbf{38}, 37-39 (2013).

\bibitem{nitrideoctave} Y. Okawachi, K. Saha, J. S. Levy, Y. H. Wen, M. Lipson, and A. L. Gaeta, ``Octave-spanning frequency comb generation in a silicon nitride chip,'' Opt. Lett. \textbf{36}, 3398-3400 (2011).
\bibitem{silicaoctave} P. Del'Haye, E. Gavartin, M. L. Gorodetsky, R. Holzwarth, and T. J. Kippenberg, ``Octave Spanning Tunable Frequency Comb from a Microresonator,'' Phys. Rev. Lett. \textbf{107}, 063901 (2011).
\bibitem{freefem} O. Pironneau, F. Hecht, A. Le Hyaric, J. Morice, ``FreeFem++,'' \url{http://www.freefem.org/}
\bibitem{singlemode} A. A. Savchenkov, I. S. Grudinin, A. B. Matsko, D. Strekalov, M. Mohageg, V. S. Ilchenko and L. Maleki, ``Morphology dependent photonic circuit elements,'' Opt. Lett., \textbf{31}, 1313 (2006).
\bibitem{sensor} I. S. Grudinin and N. Yu, ``Finite-element modeling of coupled optical microdisk resonators for displacement sensing,'' J. Opt. Soc. Am. B \textbf{29}, 3010--3014 (2012).
\bibitem{geometric}  M. L. Gorodetsky and A. E. Fomin, ``Geometrical Theory of Whispering-Gallery Modes,'' IEEE. J. Sel. Top. Quant. Electron. \textbf{12}, 33 (2006).
\bibitem{normalgvdcomb} A. B. Matsko, A. A. Savchenkov, and L. Maleki, ``Normal group-velocity dispersion Kerr frequency comb,'' Opt. Lett. \textbf{37}, 43-45 (2012).
\bibitem{diodespectroscopy} P. Del’Haye, O. Arcizet, M. L. Gorodetsky, R. Holzwarth, T. J. Kippenberg, ``Frequency comb assisted diode laser spectroscopy for measurement of microcavity dispersion,'' Nature Photonics \textbf{3}, 529 (2009).
\bibitem{supercontinuum} C. M. B. Cordeiro, W. J. Wadsworth, T. A. Birks, and P. St. J. Russell, ``Engineering the dispersion of tapered fibers for supercontinuum generation with a 1064 nm pump laser,'' Opt. Lett. \textbf{30}, 1980-1982 (2005).
\bibitem{kipraman} T. J. Kippenberg, S. M. Spillane, and K. J. Vahala, ``Kerr-Nonlinearity Optical Parametric Oscillation in an Ultrahigh-Q Toroid Microcavity,'' Phys. Rev. Lett. \textbf{93}, 083904 (2004). 
\bibitem{ramanmgf2} A. Perakis, E. Sarantopoulou, Y. S. Raptis, and C. Raptis, ``Temperature dependence of Raman scattering and anharmonicity study of MgF$_2$,'' Phys. Rev. B \textbf{59}, 775-782 (1999).

\bibitem{modelockraman} W. Liang, V. S. Ilchenko, A. A. Savchenkov, A. B. Matsko, D. Seidel and L. Maleki, ``Passively Mode-Locked Raman Laser,'' Phys. Rev. Lett. \textbf{105}, 143903 (2010).

\bibitem {smcomb} I. S. Grudinin, L. Baumgartel, and N. Yu, ``Frequency comb from a microresonator with engineered spectrum,'' Opt. Express \textbf{20}, 6604-6609 (2012). 





\end{thebibliography}
\end{document}